\documentclass[twocolumn,aps, pre]{revtex4-2}
\usepackage{float}
\usepackage{amsmath}
\usepackage{amssymb}
\usepackage{graphicx}
\usepackage[dvipsnames]{xcolor}
\usepackage{comment}
\usepackage{subfig}
\usepackage[utf8]{inputenc}
\usepackage{hyperref}
\usepackage{epstopdf}
\usepackage{siunitx}
\usepackage{ragged2e}
\usepackage{xcolor}

\usepackage{graphicx}
\usepackage{ragged2e}
\usepackage[usenames,dvipsnames]{xcolor}

\graphicspath{{./Figures/}}

\newcommand{\lpl}{\ell_\text{p}/L}
\newcommand{\lp}{\ell_\text{p}}
\newcommand{\bhat}[1]{\hat{\mathbf{#1}}}
\newcommand{\KT}{k_\text{B}T}
\begin{document}

\title{Confinement and Activity-Driven Dynamics of Semiflexible Polymers in Motility Assays}

\author{Sandip Roy}
\email[]{mp16001@iisermohali.ac.in}
\author{Abhishek Chaudhuri}
\email[]{abhishek@iisermohali.ac.in}
\author{Anil Kumar Dasanna}
\email[]{adasanna@iisermohali.ac.in}
\affiliation{Indian Institute of Science Education and Research Mohali,
Knowledge City, Sector 81, SAS Nagar 140306, Punjab, India}
\date{\today}%

\begin{abstract}
We investigate the nonequilibrium dynamics of semiflexible polymers driven by motor proteins (MPs) in two-dimensional motility assays under harmonic confinement. Using a coarse-grained agent-based model that incorporates stochastic motor attachment, detachment, and force generation, we study how activity, filament rigidity, and confinement interact to control polymer behavior. We construct dynamical behavior maps as a function of P\'eclet number, motor processivity, and trap strength. We find a two-state transition from a trapped to a free polymer, with an intermediate coexistence region. We obtain a scaling relation for the critical P\'eclet, which is supported by simulation data across a range of parameters. Polymer flexibility strongly influences confinement: flexible filaments are more easily trapped, while increasing rigidity destabilizes confinement. Processivity of MPs can also induce a change in the effective rigidity of the polymer and, therefore, influence confinement by the trap. Under moderate confinement and activity, we observe the emergence of stable spiral conformations.  The center of mass dynamics is analyzed through the mean square displacement, showing diffusive, ballistic, and diffusive regimes that depend on the trap strength and activity. Additionally, time series analysis of the excess kurtosis shows the variation of the non-Gaussian fluctuations with trap strength and activity. Our results provide a minimal physical framework to understand the dynamic organization of active filaments under confinement, with relevance to in vitro motility assays, cytoskeletal filament manipulation by optical traps, and synthetic active polymer systems. 
\end{abstract}

\maketitle
\section{INTRODUCTION}

The cytoskeleton plays a crucial role in maintaining cellular structure and dynamics. It comprises a complex, dynamic network of semiflexible filaments—actin, microtubules, and intermediate filaments—whose behavior is coordinated by motor proteins. These filaments possess persistence lengths comparable to or exceeding their contour lengths, placing them in a unique physical regime where elasticity, activity, and confinement interact in nontrivial ways. A widely studied experimental realization of this interplay is the in vitro motility assay, in which cytoskeletal filaments are propelled by surface-bound motor proteins, often within confined geometries~\cite{harada1988direct,howard1989movement,bourdieu1995spiral}.

Motility assays have been extensively explored through theory and simulation, revealing a broad spectrum of dynamical behaviors. These include spiral formation, rigidity modulation via motor attachment/detachment, crossover dynamics in center-of-mass motion, reentrant transitions between extended and coiled configurations, and collective phenomena such as gliding and swirling~\cite{duke1995gliding,scholey1993motility,chaudhuri2016forced,shee2021semiflexible,gupta2019morphological,uyeda1990myosin}. Spiral conformations have also been reported in polar active polymers driven along their tangent direction~\cite{isele2015self}. More recently, filament motion fixed at one end of the motor bed has been shown to induce beating and rotation reminiscent of ciliary motion~\cite{chelakkot2021synchronized,tiwari2020periodic,anand2019beating}.

While these advances have improved our understanding of active polymers in unconfined or bulk environments, the role of confinement—arising in cellular boundaries, vesicles, or microfluidic traps—remains comparatively underexplored, especially for semiflexible polymers subject to spatially distributed, stochastic driving forces. Confinement introduces new physical constraints and emergent behaviors. For example, boundary-induced alignment, spiral stabilization, and shape instabilities have been observed in both simulations and experiments involving confined active filaments and microswimmers~\cite{dauchot2019dynamics,malakar2020steady,chaudhuri2021active,hennes2014self,santra2021direction,wexler2020dynamics}. In porous media, where confinement arises from complex, irregular geometries rather than smooth walls, these effects are even more pronounced~\cite{theeyancheri2023active,saintillan2023dispersion,perez2021impact,theeyancheri2022migration} In motility assays, these confinement effects are amplified as filaments continuously interact with both motors and boundaries, often giving rise to confinement-stabilized dynamic states.

Moreover, experimental techniques such as optical traps (optical tweezers) have proven versatile in probing motility assays and biomolecular systems. These tools have enabled measurements of ATPase dynamics~\cite{bustamante2020single}, membrane fluctuations~\cite{merz2000pilus,mishra2016trapping} and sperm motility~\cite{dupuis1997actin}. Trapping a polymer within a region smaller than its radius of gyration also offers insight into chromosomal compaction and DNA organization~\cite{mateos2009spatially}. While equilibrium properties of confined polymers have been understood using scaling arguments~\cite{brochard1977dynamics,cordeiro1997shape}, nonequilibrium dynamics have garnered increasing attention in contexts such as polymer translocation~\cite{lubensky1999driven,luo2007influence,cohen2012stochastic,kumar2018sequencing,upadhyay2024homopolymer}, diffusion through networks, and DNA packaging in viral capsids~\cite{florin2002assembly,rahmani2007dynamics,upadhyay2024packing}.

Despite this progress, the nonequilibrium physics of semiflexible polymers subject to localized active forces and confinement remains poorly understood, particularly in regimes where activity, shape deformation, and boundaries compete to determine dynamics. Recent studies on motility assay setups have incorporated load-dependent extension and detachment rates for motor proteins (MPs), in line with experimental observations. Prior work using this framework has demonstrated that this dual dependence significantly influences the conformational and dynamical behavior of semiflexible filaments~\cite{chaudhuri2016forced,gupta2019morphological,shee2021semiflexible}. Notably, the filament exhibits a first-order transition from an open chain to a spiral conformation, accompanied by a reentrant behavior with respect to both the active extension and the motor turnover rate, defined as the ratio of attachment to detachment rates. Understanding how confinement modifies the transport and morphology of such filaments is crucial both for biological relevance and for synthetic active systems. In this work, we introduce a minimal two-dimensional model of a semiflexible polymer subject to random, localized active forces from motor proteins, within a circular trap.  

Many of the biologically relevant scenarios involving polymer confinement mentioned earlier, such as those related to cellular boundaries or DNA packaging within viral capsids, typically involve rigid boundary constraints. In contrast, our theoretical model focuses on soft confinement, inspired by optical trapping techniques such as those employed in the experiments of Simmons et al.~\cite{warrick1993vitro}. In their setup, optical tweezers were used to apply harmonic restoring forces to beads attached to actin filaments interacting with motor-protein-coated surfaces. Although their trap acted on a bead rather than the entire filament, the bead size ($0.5-2$ $\mu$m) was comparable to the filament length ($2-5$ $\mu$m), resulting in effective confinement of the whole polymer. While our model applies harmonic confinement uniformly to each monomer for simplicity and control, it captures the essential physical idea of competition between soft confinement and active propulsion. Our system presents an idealized framework that isolates and systematically explores the interplay of activity, flexibility, and confinement.



The polymer is modeled as a bead-spring chain with bending rigidity and excluded volume, while activity arises from stochastically applied persistent tangential forces of attached MPs distributed along the filament. The polymer becomes confined upon entering the trap region, allowing us to probe the interplay between activity, shape deformation, and boundary effects.
By varying filament stiffness, motor processivity, and confinement size, we construct a dynamical phase diagram that reveals several novel features. These include a trapping transition, where increased rigidity or activity leads to non-monotonic transport behavior, a confined-stabilized spiral regime, where deformation is locked by the boundary, and transitions between ballistic, diffusive, and localized states. To characterize these phases, we analyze mean squared displacement (MSD), trapping time, shape fluctuations, kurtosis of positional distributions, and tangent-tangent correlations. Our results provide physical insight into how semiflexibility and confinement shape the dynamics of active polymers, with implications for intracellular organization, synthetic active materials, and the design of shape-sensitive trapping platforms.

\begin{figure}[htb!]
\includegraphics[width=1.0\columnwidth]{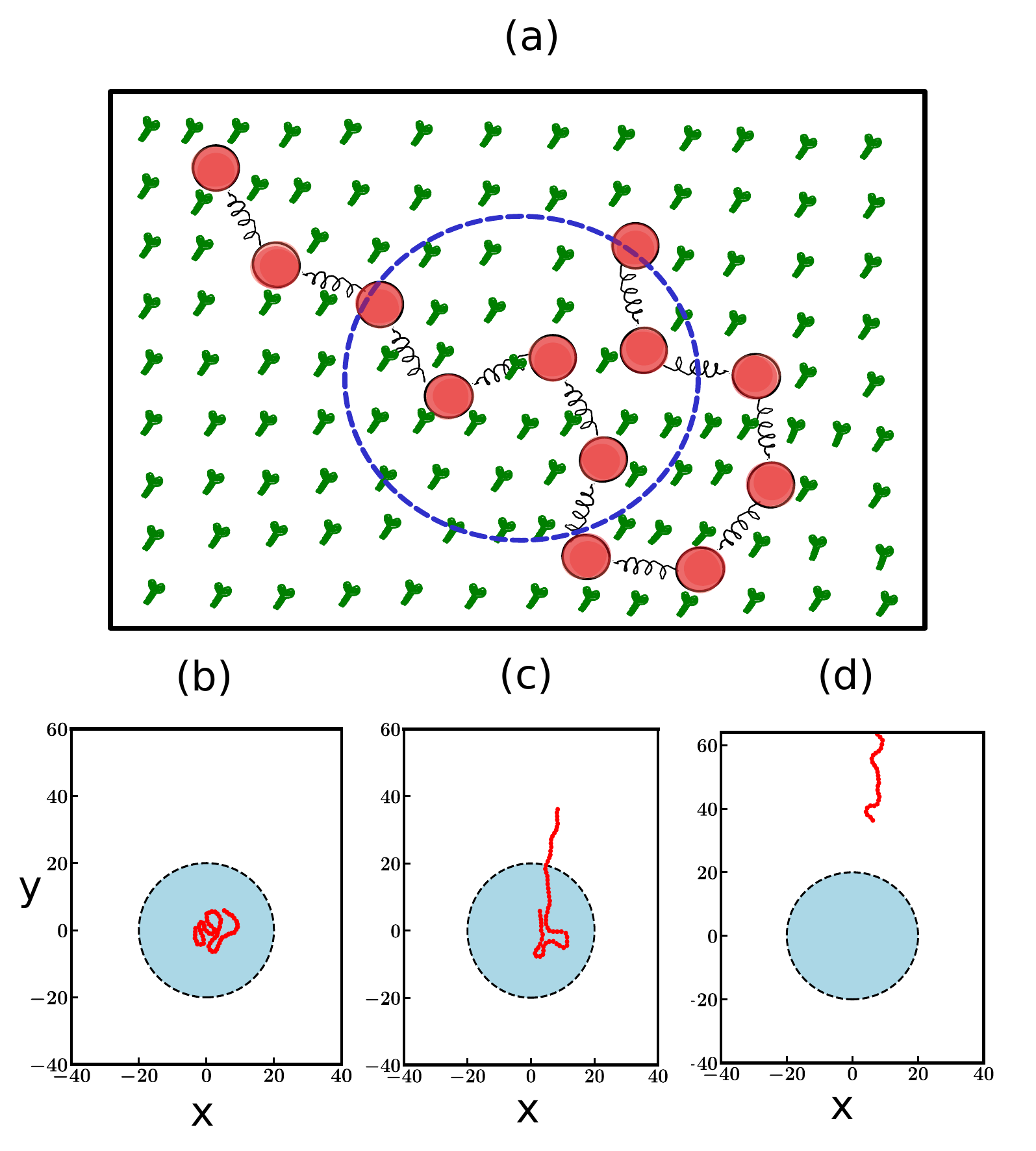}
\caption{\justifying (a) A schematic showing a semiflexible polymer (red) interacting with a grid of motor proteins (green) on a 2D substrate. A harmonic trap (blue circle) defines the confinement region.
(b–d) Simulation snapshots at increasing times for $l_p/L = 0.3, \Omega = 0.5$, and $Pe = 9.9\times 10^4$, showing transitions from a fully trapped polymer (b), to a partially trapped state (c), and finally to a free state (d).
This sequence illustrates how activity enables escape from confinement.
}
\label{fig1}
\end{figure}

\section{Model and Simulation}

We investigate the dynamics of a semiflexible polymer driven by motor proteins (MPs) in a two-dimensional motility assay in the presence of a harmonic trap. As mentioned before, the system in the presence of the MP bed has been well characterized in ~\cite{gupta2019morphological, shee2021semiflexible}. Here, we briefly discuss the model with the added trapping potential. 

The polymer is modeled as an extensible chain of $N$ monomers connected by harmonic springs and subject to bending rigidity and excluded volume interactions. Each monomer is represented by a bead with position vector $\vec{r}_i$. The bond vectors $\vec{b}_i = \vec{r}_{i+1} - \vec{r}_i$ define the local tangents $\hat{t}_i = \vec{b}_i / |\vec{b}_i|$. The total energy of the polymer consists of three contributions:

\begin{enumerate}
    \item {\textbf{Stretching energy}}:
    \begin{equation}
        {\cal E}_s = \sum_{i=1}^{N-1} \frac{K_{\mathrm{s}}}{2} \left\| \vec{b}_i - r_0 \hat{t}_i \right\|^2 
    \end{equation}
    where $K_\textrm{s}$ is the bond stiffness and $r_0$ is the equilibrium bond length.

    \item \textbf{Bending energy}:
    \begin{equation}
        {\cal E}_b = \sum_{i=1}^{N-2} \frac{\kappa}{2r_0} \left\| \hat{t}_{i+1} - \hat{t}_i \right\|^2
    \end{equation}
    where $\kappa$ is the bending rigidity.

    \item \textbf{Excluded volume interaction}: Implemented via a short-range Weeks–Chandler–Andersen (WCA) potential between non-bonded beads $i$ and $j$:
    \begin{equation}
        {\cal E}_{\text{WCA}} = 
        \begin{cases}
            4\epsilon \left[ \left(\dfrac{\sigma}{r_{ij}}\right)^{12} - \left(\dfrac{\sigma}{r_{ij}}\right)^6 \right] + \dfrac{1}{4}, & r_{ij} < 2^{1/6} \sigma \\
            0, & \text{otherwise}
        \end{cases}
    \end{equation}
\end{enumerate}

We perform molecular dynamics simulations in the presence of a Langevin heat bath, at constant temperature $k_BT = \epsilon$, unit mass $m = 1$, and isotropic damping $\gamma = 1/\tau_0$, where $\tau_0 = \sigma \sqrt{m/\epsilon}$. Simulations are performed in a finite square domain of length $L_b = 128\sigma$ with periodic boundary conditions.

\subsection{Motor Protein Dynamics}

Motor proteins are immobilized at fixed substrate positions on a square lattice of density $\rho$. Their tails are bound to the substrate, while their heads stochastically attach to nearby polymer segments within a capture radius $r_c$, at rate $\omega_{\text{on}}$ governed by a Poisson process. Once attached, the head steps actively along the polymer, mimicking the behavior of MPs such as kinesin.

The force generated by an actively stepping motor is
\begin{equation}
    \vec{f}_{\ell} = -k_m \Delta \vec{r}
\end{equation}
where $k_m$ is the linker stiffness and $\Delta \vec{r}$ is the stalk extension, connecting motor's point of contact on the polymer to immobilized position of the motor. This load is distributed among bonded monomers using a lever rule. 
The stepping velocity along the filament is load-dependent:
\begin{equation}
    v_{t}^{a}(f_t) = \frac{v_0}{1 + d_0 \exp(f_t / f_s)}
\end{equation}
where $f_t = -\vec{f}_{\ell} \cdot \hat{t}$ is the tangential load, $v_0$ is the unloaded velocity, $f_s$ is the stall force, and $d_0$ tunes force sensitivity.
The detachment rate of the motor is also load-dependent~\cite{bell1978models}:
\begin{equation}
    \omega_{\text{off}} = \omega_0 \exp(f_{\ell} / f_d)
\end{equation}
with bare rate $\omega_0$ and characteristic force $f_d$.
The net motor processivity is then given by
\begin{equation}
    \Omega(f_{\ell}) = \frac{\omega_{\text{on}}}{\omega_{\text{on}} + \omega_0 \exp(f_{\ell} / f_d)}.
\end{equation}

\subsection{Trapping}
As the polymer glides on the bed of motor proteins, it encounters a circular trap which is harmonic. The trap of strength $K_{trap}\geq 0$, centered at $\mathbf{r}_\text{o}$, is felt by any monomer within a circular region of radius $\mathcal{R}$. If the position of $i$th monomer is $\mathbf{r}_i$ then the potential is:
\begin{equation}
V_{\text{trap}} = 
\begin{cases}
\frac{1}{2} K_{\text{trap}} \left(\mathbf{r}_i - \mathbf{r}_{\text{o}}\right)^2, & \text{if } |\mathbf{r}_i - \mathbf{r}_{\text{o}}| < \mathcal{R} \\
0, & \text{otherwise}
\end{cases}
\end{equation}
 
Note that only the polymer beads experience the trap potential. The motor beads are unaffected by it. 

\subsection{Simulation Parameters and Scaling}

We use $N = 64$, $r_0 = \sigma = 1$, $K_\textrm{s} = 100~\KT/\sigma^2$, $L = (N - 1)~\sigma = 63~\sigma$ and ${\cal R} = 20~\sigma$. Our polymer and trap size already indicate that we focus on short polymers rather on larger polymers. The persistence length $\ell_p$ is related to bending rigidity as $\ell_p = 2\kappa/\KT$ in two dimensions and it is fixed at $\ell_p/L = 0.3$ unless otherwise stated, placing the filament in the semiflexible regime~\cite{shee2021semiflexible}. Activity parameters are set as $f_s = 2~k_BT/\sigma$, $f_d = f_s$, and motor density $\rho = 3.8/\sigma^2$. The motor stiffness is $k_m = K_\textrm{s}$. Activity is quantified using the dimensionless Péclet number: $\text{Pe} = {v_0 L^2}/{D \sigma}$, where $\quad D = {k_BT}/{\gamma}$. 
The time unit is defined as: $\tau = {L^3 \gamma}/{4 \sigma k_BT}$. Simulations are run up to $2 \times 10^8$ steps with timestep $\delta t \approx 1.6 \times 10^{-8} \tau$. Initial $10^{8}$ steps are discarded to ensure steady-state sampling. The detailed list of the various parameters and their values are given in Table~\ref{table}. The relevance of the values in real systems is discussed later.


\begin{table}[h!]
\small
  \caption{Different parameters and their numerical values used in the simulation:}
  \label{tbl:example1}
  \begin{tabular*}{0.48\textwidth}{@{\extracolsep{\fill}}lll}
    \hline
    Parameters & Definition & Values \\
    \hline
    m & Mass of each bead & 1\\
    N & Number of polymer beads & 64 \\
    $r_0$ & Bond length  & 1\\
    $k_BT$ & Energy scale  & 1\\
    $r_c$ & Capture radius & $0.5~\sigma$ \\
    $K_\textrm{s}$ & Spring constant of filament &100 $k_BT/\sigma^2$ \\
    $\rho$ &Density of MP &$3.8~\sigma^{-2}$\\
    $f_d$ & Detachment force &2 $k_BT/\sigma$\\
    $f_s$ & Stall force &2 $k_BT/\sigma$\\
    $d_0$ & Force sensitivity parameter &0.012\\
    $\zeta_{MP}$ & Frictional coeffecient of MP &$0.1~\zeta$\\
    $k_m$ & Elastic coeffecient of MP &100 $k_BT/\sigma^2$ \\  
    $\mathcal{R}$ & Radius of Confinement &$20~\sigma$\\
    $\Omega$ & Bare processivity rate &$0.1-0.9$\\
    $K_{trap}$ & Trap strength &  $0.03-0.1~k_BT/\sigma^2$\\
    $Pe$ &P\'eclet number& $1.9\times10^4-29.7\times10^4$\\
    \hline
  \end{tabular*}
  \label{table}
\end{table}

\begin{figure*}[htb!]
\includegraphics[width=\textwidth]{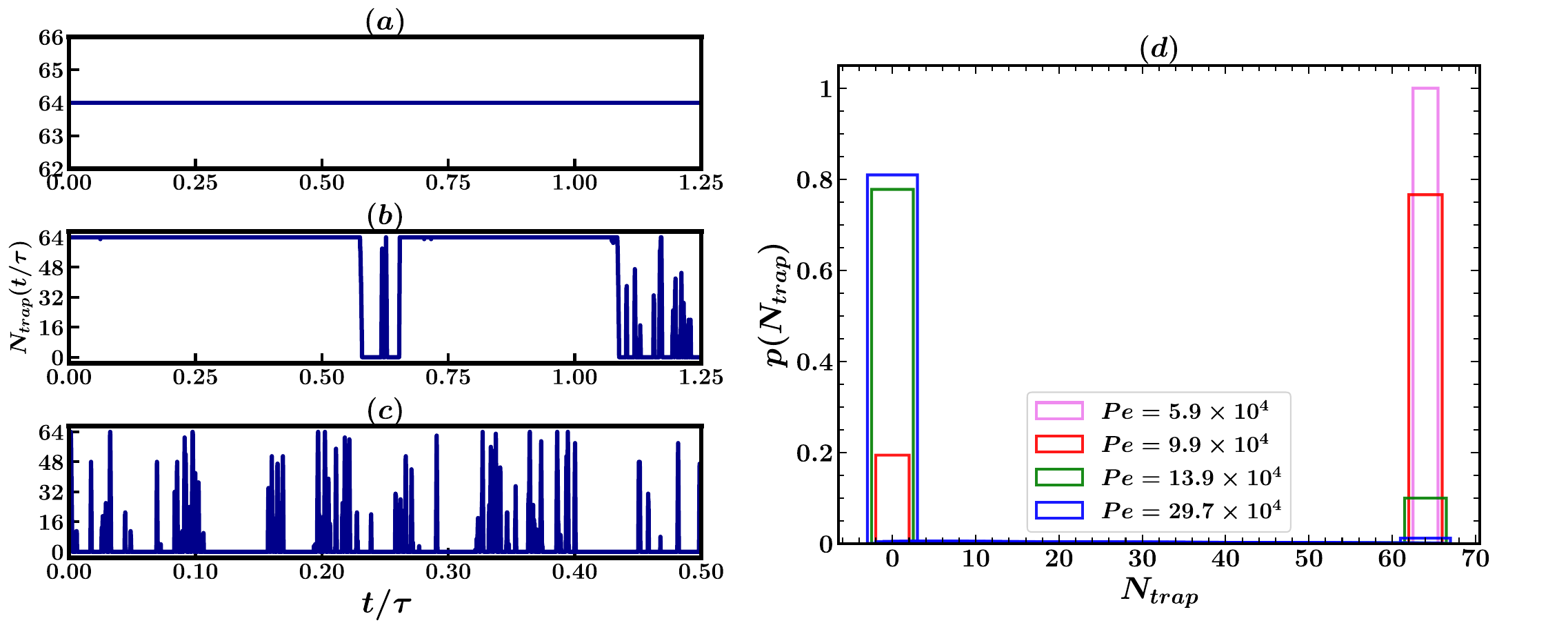}
\caption{\justifying 
(a–c) Time series of the number of monomers $N_{trap}$ within the trap for increasing P\'eclet numbers, at fixed $\Omega = 0.5$ and \( K_{\text{trap}} = 0.1 \KT/\sigma^2 \). (d) Probability distribution \( p(N_{\text{trap}}) \) for three P\'eclet numbers at \( K_{\text{trap}} = 0.06 \KT/\sigma^2 \). At low Pe, the polymer remains confined; intermediate Pe shows bimodal behavior; high Pe favors escape. Different bar widths are used for visual clarity.
}
\label{fig2}
\end{figure*}

\section{Results}
We initialize our simulations with the polymer fully enclosed within a circular harmonic trap. In the absence of the trap, the polymer dynamics is governed solely by three parameters: the P\'eclet number ($\mathrm{Pe}$), the motor processivity ($\Omega$), and the polymer stiffness, quantified by the persistence length to contour length ratio ($\ell_p/L$). The presence of the trap introduces additional constraints, making the dynamics dependent also on the trap stiffness ($K_{\mathrm{trap}}$) and the fraction of the area occupied by the trap.

Our simulations reveal that the system predominantly exhibits two distinct dynamical states: the polymer either remains completely confined within the trap or escapes into a free state (see Fig.~\ref{fig1}(b) and Fig.~\ref{fig1}(d)). Although the polymer may intermittently re-enter the trap region in the free state, its activity ensures that it escapes the trap rapidly, preventing sustained confinement. Between these two limiting behaviors, we also observe metastable intermediate states, where the polymer remains transiently trapped for extended periods before eventually escaping. The occurrence of these different states depends sensitively on the trap strength, motor activity, and polymer rigidity.

To distinguish between these states quantitatively, we define $N_{\mathrm{trap}}$ as the number of monomers inside the trap region. We examine the time evolution of $N_{\mathrm{trap}}$ for three different values of $\mathrm{Pe}$, keeping the bare motor processivity fixed at $\Omega = 0.5$ and the polymer persistence ratio at $\ell_p/L = 0.3$ (see Fig.~\ref{fig2}(a)-(c)). The trap strength is chosen such that the polymer remains confined at $\mathrm{Pe} = 0$, with $K_{\mathrm{trap}} = 0.1\,k_BT/\sigma^2$. At zero activity ($\mathrm{Pe} = 0$), the polymer is stably trapped, and $N_{\mathrm{trap}}$ shows no variations (Fig.~\ref{fig2}(a)). As activity increases ($\mathrm{Pe} = 17.8 \times 10^4$ and $29.7 \times 10^4$), transient excursions outside the trap become more frequent and pronounced (Fig.~\ref{fig2}(b)-(c)), indicating the destabilization of the confined state.

Fig.~\ref{fig2}(d) shows the probability distribution $P(N_{\mathrm{trap}})$ for different $\mathrm{Pe}$ values at fixed trap strength. Fully trapped polymers correspond to $P(N_{\mathrm{trap}} = 64) \approx 1$, whereas free polymers yield $P(N_{\mathrm{trap}} = 0) \approx 0.8$. In contrast, intermediate states would show broader distributions with reduced peak probabilities, reflecting their metastable nature. As observed in Fig.~\ref{fig2}(d), for low activity, there is a single peak at $N_{\mathrm{trap}} = N$. For moderate activity, a secondary peak at $N_{\mathrm{trap}} = 0$ emerges, indicating intermittent escapes. At high activity, the distribution shifts further, reflecting the dominance of the free polymer state.

\begin{figure*}[htb!]
\includegraphics[width=0.8\textwidth]{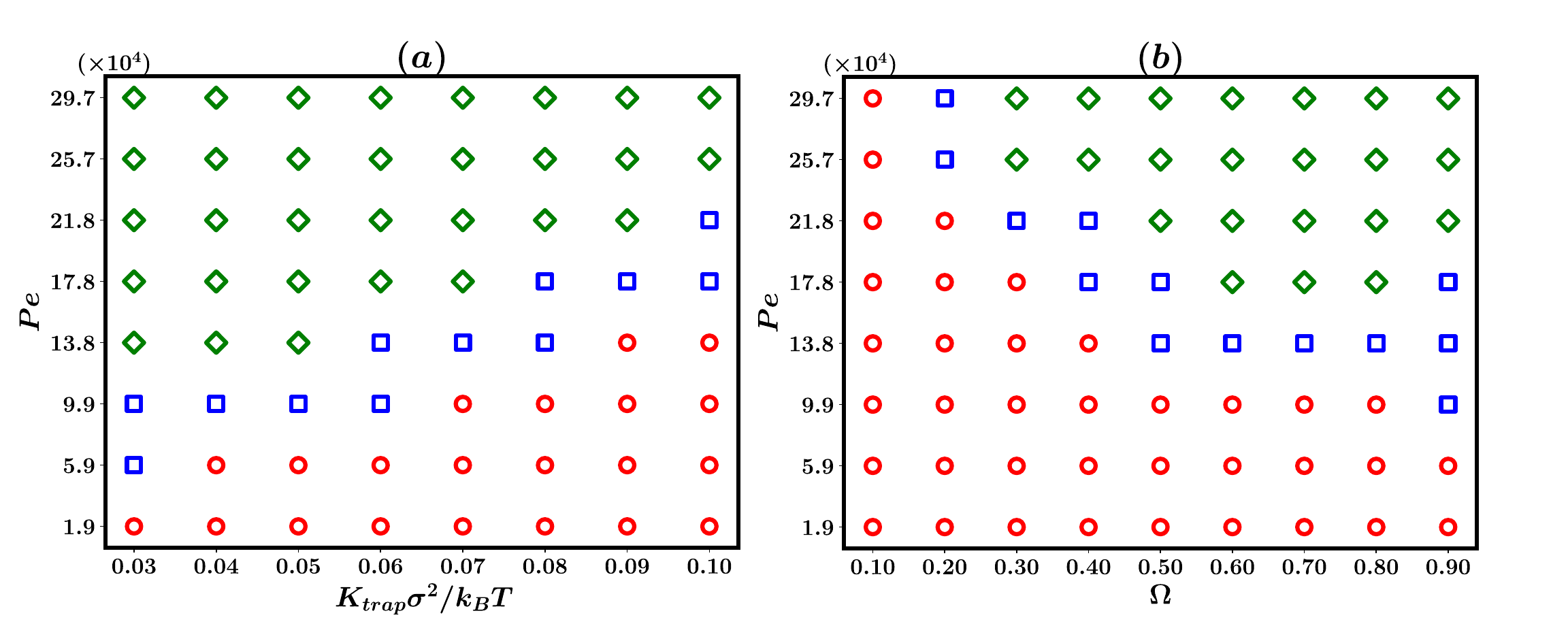}
\caption{{Dynamical behavior maps (a) in the $K_{\text{trap}}$-P\'eclet plane at $\Omega = 0.5$ and (b) in the $\Omega$-P\'eclet plane at fixed 
$K_{\text{trap}} = 0.08 ~\KT/\sigma^2$.}
Green diamonds ({\color{OliveGreen}\Large $\diamond$}) represent free polymer states, red circles ({\color{Mahogany}\Large $\circ$}) denote trapped states, and blue squares ({\color{RoyalBlue}$\square$}) indicate coexistence or metastable states. The diagrams highlight how activity and motor processivity control confinement transitions.
}
\label{state}
\end{figure*}

\subsection{Dynamical Regimes}

We construct dynamical behavior maps as a function of various parameters (Fig.~\ref{state}(a) and (b)). The maps classifies the observed behaviors into three broad regimes - trapped, free, and intermediate - based on the escape probability inferred from the time-averaged distribution of trapped monomers. While not true thermodynamic states, these regimes capture qualitatively distinct dynamical responses over finite simulation times. These maps are constructed based on the probability histogram of $N_{\text{trap}}$, computed over long simulation runs for various values of $Pe$, $\Omega$, and $K_{\text{trap}}$. Specifically, we analyze the \emph{maximum height} of this histogram, which reflects the likelihood of the polymer occupying a specific trapped state.



In Fig.~\ref{state}(a), we show the dynamical regimes in the $K_{\text{trap}}$-P\'eclet plane. The transition from the trapped to the free state appears monotonic at fixed processivity. This behavior is intuitively expected: a higher trap stiffness requires a larger active drive to overcome confinement. Consequently, at lower trap strengths, the transition to the free state occurs at smaller $\mathrm{Pe}$, while at higher $K_{\text{trap}}$, the transition requires progressively larger $\mathrm{Pe}$.

To investigate the role of motor protein (MP) binding dynamics, we also construct a similar diagram varying $\mathrm{Pe}$ and $\Omega$, at a fixed trap strength $K_{\text{trap}} = 0.08\,k_BT/\sigma^2$ (Fig.~\ref{state}(b)). At lower processivity, the polymer predominantly remains trapped, and even increasing $\mathrm{Pe}$ does not lead to escape. At higher processivity, however, the polymer transitions more readily into the free state with increasing $\mathrm{Pe}$.

This finding is particularly significant as it hints towards the broader role played by the load dependent attachment/detachment kinetics of MPs in determining the statics and dynamics of polymers in the presence of activity. This would not be possible in studies which consider the polymers to be made up of active monomers with a constant velocity in the tangential direction or introduce activity via an active noise term \cite{jiang2014motion,isele2015self,eisenstecken2016conformational,winkler2017active,man2019morphological,winkler2020physics}\cite{chelakkot2014flagellar,ghosh2014dynamics,shin2015facilitation,de2017spontaneous,duman2018collective,prathyusha2018dynamically,mokhtari2019dynamics,peterson2020statistical,anand2020conformation} 

To understand the observed behavior, we fix the $Pe$ at a moderate value (say, $Pe \sim 25 \times 10^4$), and increase the processivity. At low processivity, the polymer forms stable spirals inside the trap (see Section E for more detailed discussion). As the processivity increases, the attachment rate increases, generating stronger active forces and increased fluctuations. This facilitates escape from the trap.

It is important to note that the trap radius significantly influences the behavior. Within the range of $\mathrm{Pe}$ and $\Omega$ considered, a trap much larger than the polymer size inevitably captures the polymer. In comparison, a much smaller trap consistently allows the polymer to remain unconfined. Supporting examples can be found in Figs.~\ref{Fig1} and ~\ref{Fig2} of the Supplementary Information.

\subsection{Area function dependence of escape probability}
\begin{figure}[htb!]
\includegraphics[width=\columnwidth]{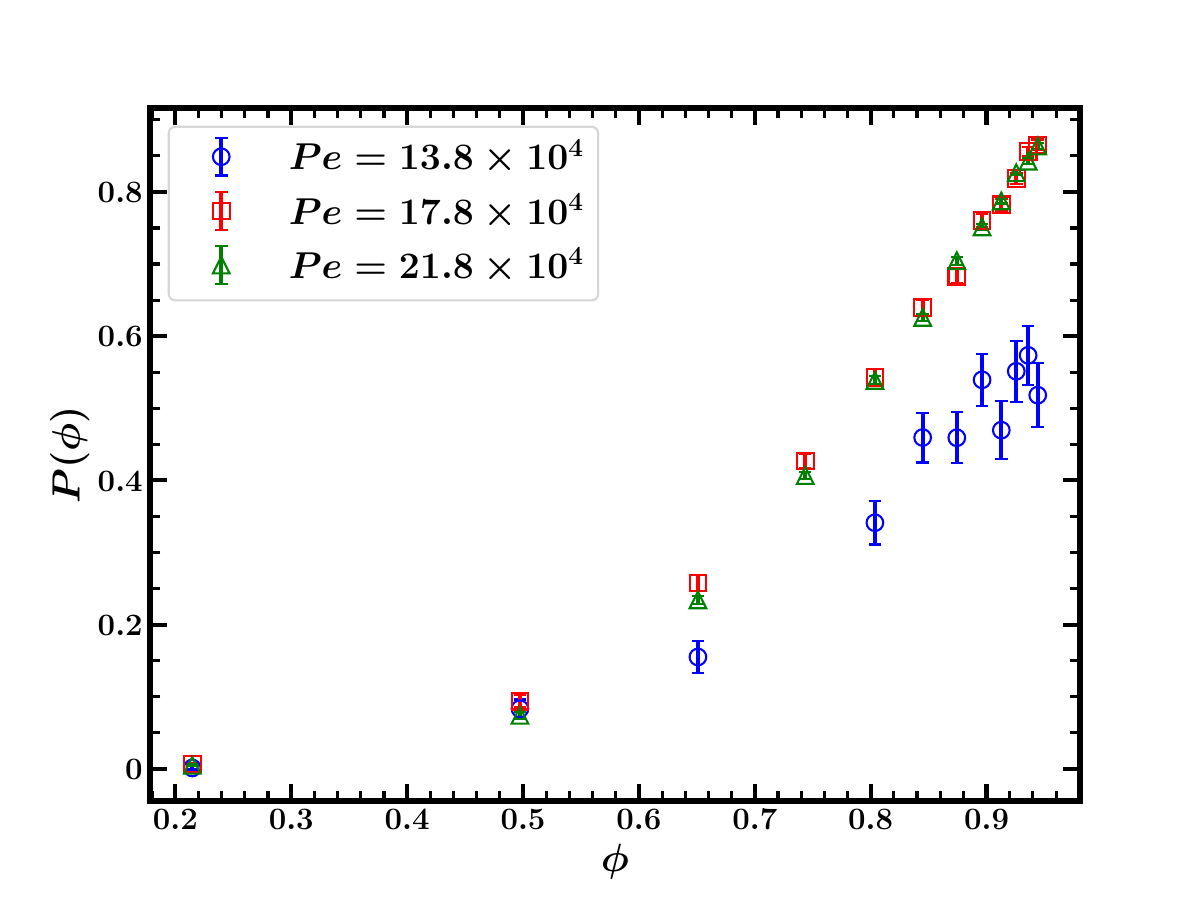}
\caption{\justifying 
Escape probability $P(\phi)$, defined as the probability the polymer remains outside the trap, plotted against area fraction $\phi$ 
for three P\'eclet numbers. Simulations use $\Omega = 0.5$ and \( K_{\text{trap}} = 0.08 k_B T / \sigma^2 \). Larger free space facilitates escape, and the threshold shifts to lower $\phi$ with increasing $Pe$. Error bars computed from simulations with different random seeds are shown in the plot.
}
\label{escape}
\end{figure}

The polymer's behavior outside the trap is influenced by the extent of free space available for exploration before returning to the confinement region. To systematically address this, we varied the area fraction $\phi = (L_b^2 - \pi \mathcal{R}^2)/L_b^2$, where $L_b$ is the simulation box length. The probability that the polymer is located outside the trap at a given area fraction is $P(\phi)$. In our default simulations, $L_b = 128\,\sigma$ and $\mathcal{R} = 20\sigma$, corresponds to an area fraction $\phi \simeq 0.92$.

Figure~\ref{escape} shows the evolution of $P(\phi)$ as a function of $\phi$ for three different Péclet numbers. The simulations are performed at a trap strength $K_{\text{trap}} = 0.08\,k_B T/\sigma^2$, with bare processivity $\Omega = 0.5$ and persistence ratio $\ell_p/L = 0.3$. The selected Péclet numbers are $\mathrm{Pe} = 13.8 \times 10^4$, $17.8 \times 10^4$, and $21.8 \times 10^4$, where the first two correspond to transiently trapped states and the last to a free state. As $\phi$ increases, the polymer spends more time outside the confinement region. The polymer remains effectively trapped for small values of $\phi$, with a negligible escape probability, i.e., $P(\phi) \approx 0$. As $\phi$ increases, the polymer gains access to larger free regions, facilitating escape from the trap. The escape probability approaches unity in the limit of very large $\phi$.

For the default area fraction ($\phi \simeq 0.92$), even though $\mathrm{Pe} = 13.8 \times 10^4$ corresponds to a transient state (see Fig.~\ref{state}(b)), the polymer exhibits a slight preference for remaining trapped but eventually escapes. This subtle competition between being trapped for longer durations and having more free space to explore leads to an oscillatory trend in the probability $P(\phi)$ at intermediate $\phi$. For higher Péclet numbers, the probability curves become smoother. 
At higher values of $\mathrm{Pe}$, the probability curves for different Péclet numbers overlap, indicating statistical similarity in the polymer’s escape behavior.

It is natural to ask how the overall system would be affected in the limit of $\phi \to 1$. In this limit, as long as the initial configuration of the polymer corresponds to a trapped state, the dynamical regimes remain qualitatively unchanged. However, if the initial polymer state is not trapped, the final dynamical state becomes determined predominantly by the Péclet number and the bare processivity $\Omega$. In the opposite limit, as $\phi \to 0$, the polymer always remains trapped.

\begin{figure}[ht]
\includegraphics[width=\columnwidth]{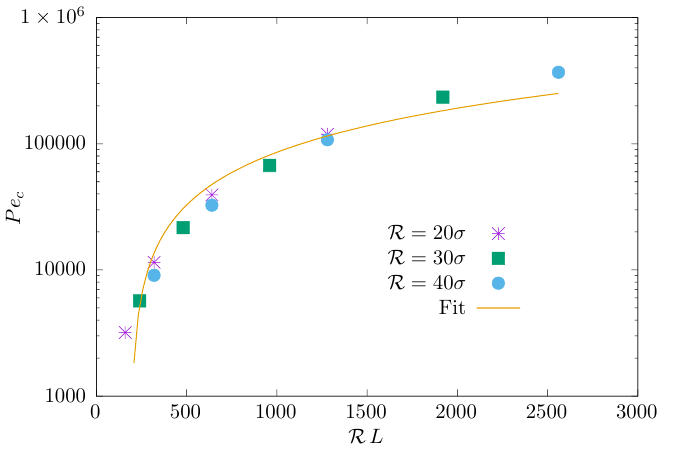}
\caption{\justifying
Critical Péclet number $Pe_c$  required for escape, plotted against the scaled parameter ${\cal R}\,L$ for various trap radii. The fit (solid line) confirms the scaling relation $Pe_c \sim K_{\text{trap}}L\,{\cal R}$, derived in Section III.C. Semilogarithmic plot shows good agreement between theory and simulation.
}
\label{scaling}
\end{figure}

\subsection{Critical P\'eclet number: A scaling analysis}
We now present a scaling argument to understand the dependence of the transition between trapped and free states on the key system parameters: polymer length \(L\), trap radius \(\mathcal{R}\), and trap stiffness \(K_{\text{trap}}\).
MP force per unit length exerted on the polymer depends on $f_l$ force exerted due to the active extension of the MPs, the linear density $\sqrt{\rho}$ and the processivity $\Omega(f_l)$. The mean active force $f_a \sim \gamma v_0$. The net active force per unit length $\sim \sqrt{\rho}\Omega(f_l)f_a$. To determine the escape threshold, we equate the net active force for a polymer of length $L$ to the trap force $K_{trap}\mathcal{R}$. Thus:
\[
\gamma \sqrt{\rho}\Omega(f_l)v_0L \sim K_{trap}\mathcal{R}
\]
which gives $v_0 \sim K_{trap}\mathcal{R}/\left(\gamma \sqrt{\rho}\Omega(f_l)L\right)$. Now, $Pe = v_0L^2/D\sigma$ with $D = k_BT/\gamma$. Therefore, the critical P\'eclet for escape:
\begin{eqnarray}
\nonumber
Pe_c &\sim \frac{\left(K_{trap}\mathcal{R}/\left(\gamma \sqrt{\rho}\Omega(f_l)L\right)\right)L^2}{k_BT\sigma/\gamma} \\
&= \frac{K_{trap}\,L\,\mathcal{R}}{\sqrt{\rho}\,\Omega(f_l)\,k_BT\,\sigma} \sim K_{trap}\,L\,\mathcal{R} 
\label{scale}
\end{eqnarray}
In Fig.~\ref{scaling}, we plot the critical $Pe$ for escape with changing polymer length $L$ and trap radius $\mathcal{R}$ according to Eq.~\ref{scale}.

It is important to note that the scaling form presented in Eq.(9) is derived under a simplified mean-field assumption where the net active force is estimated from average motor activity and linear drag, and equated to the restoring force from the harmonic trap. This approach neglects several subdominant but potentially relevant contributions, such as thermal fluctuations, reentrant excursions into the trap due to finite activity, and torque generated by spiral configurations. Moreover, the effect of polymer stiffness enters only implicitly via the assumed processivity ($\Omega(f_l)$), but does not account for shape-induced asymmetries or persistence-length-dependent escape trajectories. As a result, while the scaling captures the leading order dependence on ${\cal R}$ and $L$, it may not hold quantitatively across regimes with strong confinement, high rigidity 
$l_p \sim L$, or nontrivial trap geometries.

\begin{figure*}[htb!]
\includegraphics[width=\textwidth]{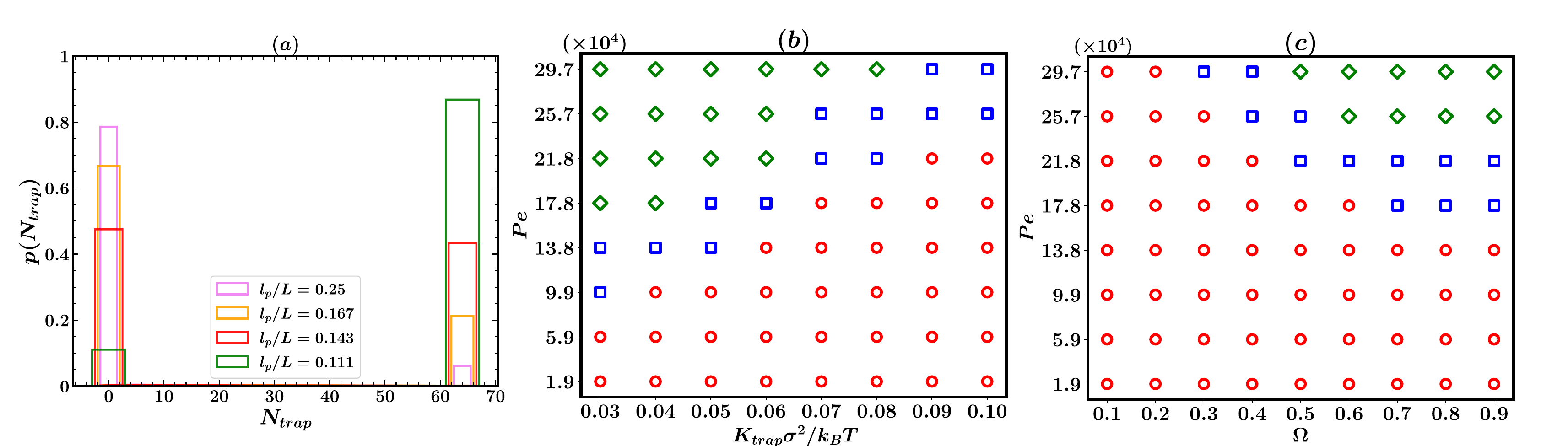}
\caption{\justifying 
(a) Distribution \( p(N_{\text{trap}}) \) for various persistence ratios \( \lpl \), at fixed \( Pe = 13.8 \times 10^4 \) and \( K_{\text{trap}} = 0.06~\KT/\sigma^2 \).
(b, c) Dynamical behavior maps for a softer polymer ($\ell_p/L = 0.1$) analogous to Fig. 3. Increased flexibility enhances trapping, requiring higher $Pe$ for escape.
}
\label{rigidity}
\end{figure*}

\begin{figure*}[htb!]
\includegraphics[width=\textwidth]{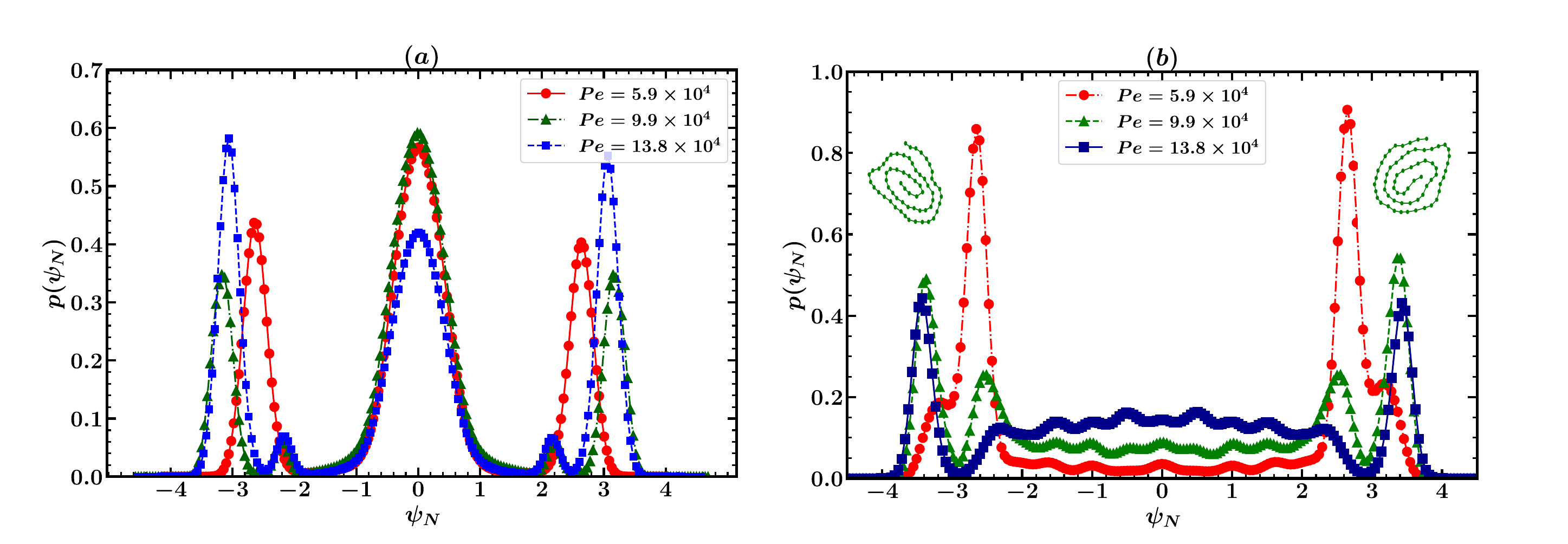}
\caption{\justifying 
(a) Steady-state distribution of turning number $\psi_N$ without confinement. (b) Same distribution with strong confinement $K_{\text{trap}} = 0.4~\KT/\sigma^2$, showing stable spiral peaks. Representative configurations, as inset snapshots, show counter-clockwise (positive $\psi_N$; right) and clockwise spirals (negative $\psi_N$; left). Trap promotes spiral stability at intermediate activity.
}
\label{fig4}
\end{figure*}

\subsection{Effect of polymer rigidity}
In Fig.~\ref{rigidity}(a), we illustrate the explicit effect of the persistence ratio, \( \lpl \), on the probability distribution of trapped monomers. The activity is kept fixed at $Pe = 13.8 \times 10^4$, with a confinement strength of $K_{\text{trap}} = 0.06~\KT/\sigma^2$ and a bare processivity rate of $\Omega = 0.5$. We systematically vary the persistence ratio $\lp/L$, transitioning from flexible to stiffer polymer conditions. Our results show that a flexible polymer is easily confined and requires higher activity to escape the trap. For $\lp/L = 0.111$, the polymer remains almost confined within the trap. For $\lp/L = 0.143$, the distribution $p(N_{\text{trap}})$ shows two peaks at $N_{\text{trap}} = 0$ and $N_{\text{trap}} = 64$, both of equal height, suggesting a metastable state between the trapped and free configurations.  As $\lp/L$ increases further to $\lp/L = 0.167$, the peak at $N_{\text{trap}} = 0$ becomes more pronounced, while the peak at $N_{\text{trap}} = 64$ diminishes, indicating reduced stability in the trapped state. Finally, for $\lpl = 0.25$, the peak at $N_{\text{trap}} = 64$ nearly vanishes, with a strong peak at $N_{\text{trap}} = 0$, signifying a progressive destabilization of the trapped state as the polymer becomes stiffer.

We have also provided a detailed maps representing the different regimes for a softer polymer $\lpl=0.1$ in Figs.~\ref{rigidity}(b-c) to compare with Fig.~\ref{state}(a-b). In both cases, the trapped regime increases significantly compared to Fig.~\ref{state}. The transition to a free state now requires a significantly larger $Pe$, i.e. a larger velocity of attached MPs. Interestingly, the qualitative behavior remained largely unchanged, namely, a nearly monotonic increase in the P\'eclet number $Pe$ with trap strength, along with a non-monotonic change in $Pe$ with the base processivity $\Omega$.

\subsection{Spiral formation stabilized by confinement}
It was shown that when the activity is non-zero i.e $Pe\neq 0$, the semiflexible polymer in the presence of motility assay \cite{gupta2019morphological} or even an active polar polymer~\cite{isele2015self} displays open chain and spiral configurations. These spirals can have clockwise and counterclockwise orientations. In the motility assay, the polymer undergoes a first order phase transition from the open chain to spiral conformation and shows a reentrant behavior in both $Pe$ and $\Omega$~\cite{shee2021semiflexible}. In this section, we quantify this behavior in the presence of the trap. 

To quantify spiral formation, we use the turning number, defined as 
$\psi_i = \frac{1}{2\pi}\sum_{j=1}^{i-1}[\phi_{j+1}-\phi_j]$
where $\phi_j$ is defined by $ \hat{t}_j= (\cos \phi_j,\sin \phi_j)$, and $\phi_{j+1}-\phi_j$ gives the angle increment between consecutive bonds~\cite{shee2021semiflexible}. Then, $\psi_N$, quantifies the number of turns the polymer undergoes across its full length. For a straight chain, $\psi_N = 0$, while for a single loop, $\psi_N = \pm 1$ for clockwise/anticlockwise turns, the magnitude of $\psi_N$ increases for multiple turns, indicating a spiral conformation. The value of $\psi_N$ depends primarily on the length of the polymer and is measured in the steady state.

In Fig.~\ref{fig4}(a), we show the distribution of $\psi_N$, $p(\psi_N)$, for three $Pe$ values without a trap ($K_{\text{trap}} = 0$). At $Pe = 5.9 \times 10^4$, a central peak near $\psi_N \approx 0$ indicates an open-chain gliding motion, with secondary peaks at $\psi_N \approx \pm 2.5$ corresponding to unstable spirals. At $Pe = 9.9 \times 10^4$, the central peak remains, but two new spiral peaks appear at $\psi_N \approx \pm 3$. As $Pe$ increases to $13.8 \times 10^4$, the central peak diminishes, signaling the instability of open-chain motion, and stable spirals form at $\psi_N \approx \pm 3$. At higher $Pe$, the spirals lose stability, marking a re-entrant transition~\cite{shee2021semiflexible}.

\begin{figure}[ht]
\includegraphics[width=\columnwidth]{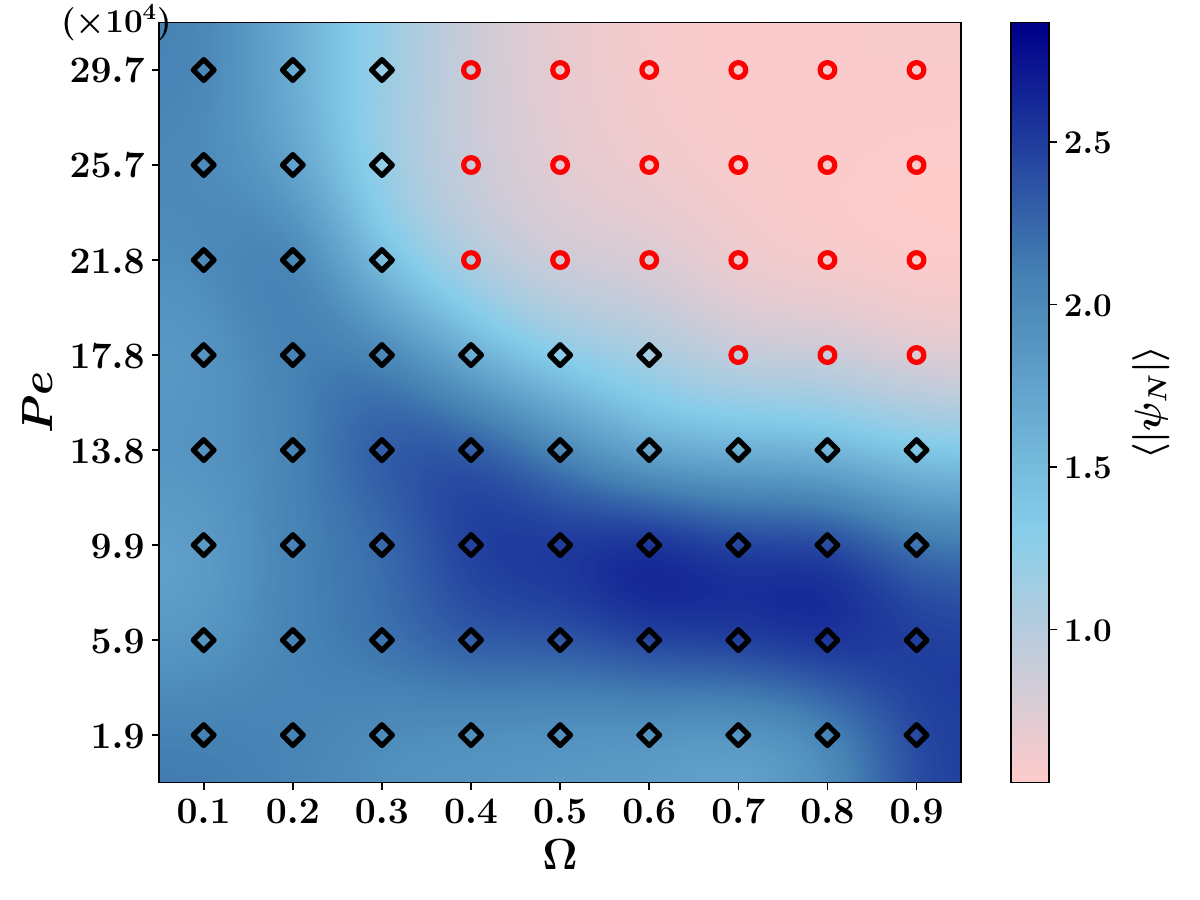}
\caption{\justifying An average absolute spiral number $\langle |\psi_N| \rangle$ as a function of activity ($Pe$) and the bare processivity rate is shown keeping $L=63\sigma,\Omega=0.5,\mathcal{R}=20\sigma$.  Red circles indicate unstable spiral or open-chain conformations, while black diamonds represent stable spirals. The color gradient reflects the magnitude of $\langle |\psi_N| \rangle$: values less than 1 are shown in reddish tones, and values greater than 1 are depicted in blue.}
\label{phaseDiag_spiral}
\end{figure}

\begin{figure*}[ht]
\includegraphics[width=\textwidth]{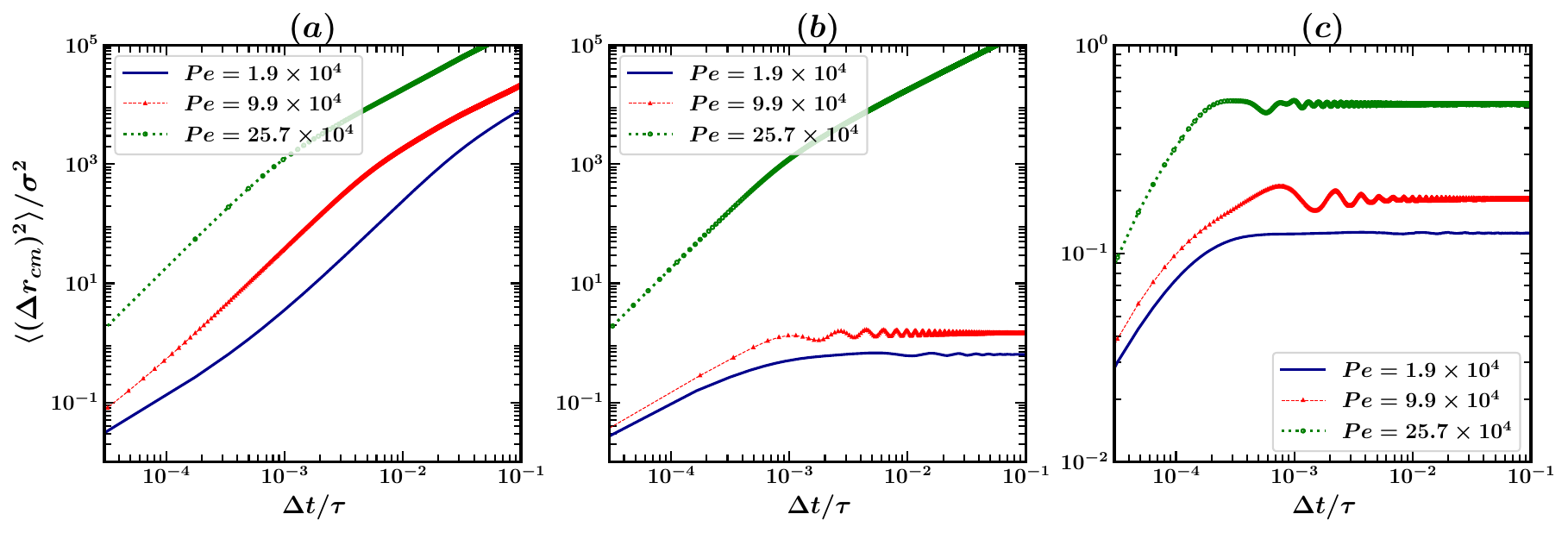}
\caption{\justifying 
MSD of the polymer’s center of mass at three Péclet numbers, for trap strengths: (a) $K_{\text{trap}} = 0$ (no trap), (b) $K_{\text{trap}} = 0.1~\KT/\sigma^2$, and (c) $K_{\text{trap}} = 0.5~\KT/\sigma^2$.
Transitions from diffusive to ballistic to saturated regimes are observed. Oscillations at intermediate Pe signal spiral formation. The bare processivity rate is kept to be $\Omega=0.5$ throughout.
}
\label{MSD}
\end{figure*}

When a harmonic trap is applied ($K_{\text{trap}} = 0.4~\KT/\sigma^2$), with $Pe$ chosen so that the polymer stays within the trap, the impact on spiral formation is shown in Fig.~\ref{fig4}(b). The gliding motion is stopped, eliminating the central peak. At $Pe = 5.9 \times 10^4$, two stable spirals emerge at $\psi_N \approx \pm 3$ with greater stability compared to $K_{trap}=0$. As $Pe$ increases to $9.9 \times 10^4$, stable spirals appear at $\psi_N \approx \pm 2.5$ and $\psi_N \approx \pm 3.5$. At $Pe = 13.8 \times 10^4$, the peak at $\psi_N \approx \pm 2.5$ disappears, leaving a single peak at $\psi_N \approx \pm 3.5$, nearly identical to the peak at $Pe = 9.9 \times 10^4$. Two representative snapshots from the simulations are also shown for positive (clockwise) and negative (counter-clockwise) turning numbers for $Pe = 9.9 \times 10^4$.

Trapping and processsivity also influence spiral formation and stabilization. In the absence of a trap, the polymer largely shows open chain conformations at low processivity. However, trap stabilizes spiral formation even at low processivity (see Supplementary Fig. \ref{Fig3}). Increasing processivity results in more motors remaining attached, generating stronger active forces and increased fluctuations. As a result, the polymer exhibits more frequent transitions between clockwise and counterclockwise spirals inside the trap, characterized by various turning numbers $\psi_N$. The presence of multiple spirals (both clockwise and counterclockwise) with different $\psi_N$ allows the polymer to adopt diverse conformations, facilitating a broader range of structural configurations. 

We extended our analysis by computing the average absolute turning number $\langle |\psi_N| \rangle$, which quantifies the degree of spiral formation across different Péclet numbers and processivity rates. This measure captures the typical number of turns in a polymer configuration, independent of the spiral’s direction. As shown in Fig.~\ref{phaseDiag_spiral}, regions where 
$\langle |\psi_N| \rangle < 1$ (marked with red circles) correspond to linear or unstable spiral configurations, while 
$\langle |\psi_N| \rangle > 1$ (black diamonds) indicates robust, stable spirals.


Thus, activity and confinement alters the effective mechanical properties of the polymer. The directed forces applied by attached motor proteins introduce persistent stresses along the filament contour. These active stresses can effectively reduce the polymer's ability to bend thermally by aligning local segments, thereby increasing the apparent stiffness at short timescales. At larger activities, frequent motor detachment and reattachment events introduce localized distortions that remodel bending. These effects become significant near circular trap boundaries, leading to the stabilization of spiral conformations. Moderate activity levels enhance the filament's tendency to curve without generating large distortions, allowing the polymer to wrap smoothly into a stable, compact spiral within the trap. 
Thus, spiral stabilization emerges from a balance between active force generation, motor-induced remodeling of local stiffness, and geometrical confinement. Such spiral shaped configurations may be observed using high-resolution fluorescence imaging in actin-gliding assays confined by microfabricated traps. 

\subsection{Dynamics of centre of mass}
\begin{figure}[htb!]
\includegraphics[width=\columnwidth]{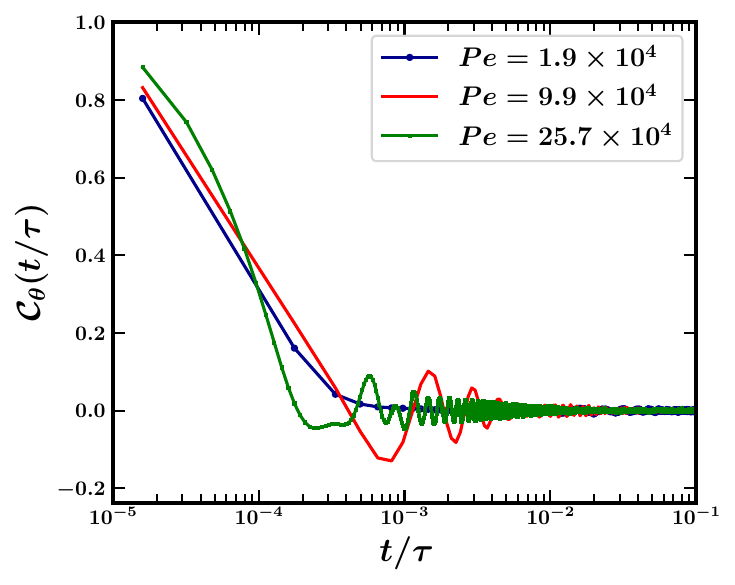}
\caption{\justifying 
Orientation autocorrelation $C_\theta(t) = \left\langle \exp^{i[\theta(t) - \theta(0)]} \right\rangle$ at $K_{\text{trap}} = 0.5\KT/\sigma^2$, $\Omega=0.5$ for three $Pe$ values. Decay rates and oscillations increase with $Pe$, indicating spiral rotation and active realignment.}
\label{cross-cor}
\end{figure}
The mean squared displacement (MSD) of the polymer's center of mass provides crucial insight into the transition between trapped and untrapped states.  The MSD is defined as 
\[
MSD(\Delta t) = \langle (\mathbf{r}_{cm}(t+\Delta t) - \mathbf{r}_{cm}(t))^2 \rangle,
\]
where \(\mathbf{r}_{cm}(t)\) represents the position of the center of mass of the polymer at time \(t\). 

In Fig.~\ref{MSD}(a), we show the $MSD$ in the absence of confinement (\(K_{trap} = 0\)). This reveals three different regimes depending on the value of \(Pe\). For \(Pe = 1.9 \times 10^4\), three behaviours are observed: (1) a diffusive regime (\(MSD \propto \Delta t/\tau)\) for \(\Delta t/\tau \lesssim 10^{-3}\), (2) a ballistic regime (\(MSD \propto (\Delta t/\tau)^2\)) for intermediate times, and (3) a return to diffusive behaviour for \(\Delta t/\tau \gtrsim 10^{-1}\). Increasing the activity parameter shifts the diffusive regime earlier, with the first diffusive-ballistic crossover occurring at $\Delta t/\tau \lesssim 10^{-4}$ for $Pe = 9.9\times 10^4$. 
An analogous sequence comprising diffusive, ballistic and then diffusive behaviours have been observed in the case of individual active Brownian particles~\cite{howse2007self,martens2012probability,bechinger2016active}.

\begin{figure*}[htb!]
\includegraphics[width=0.9\textwidth]{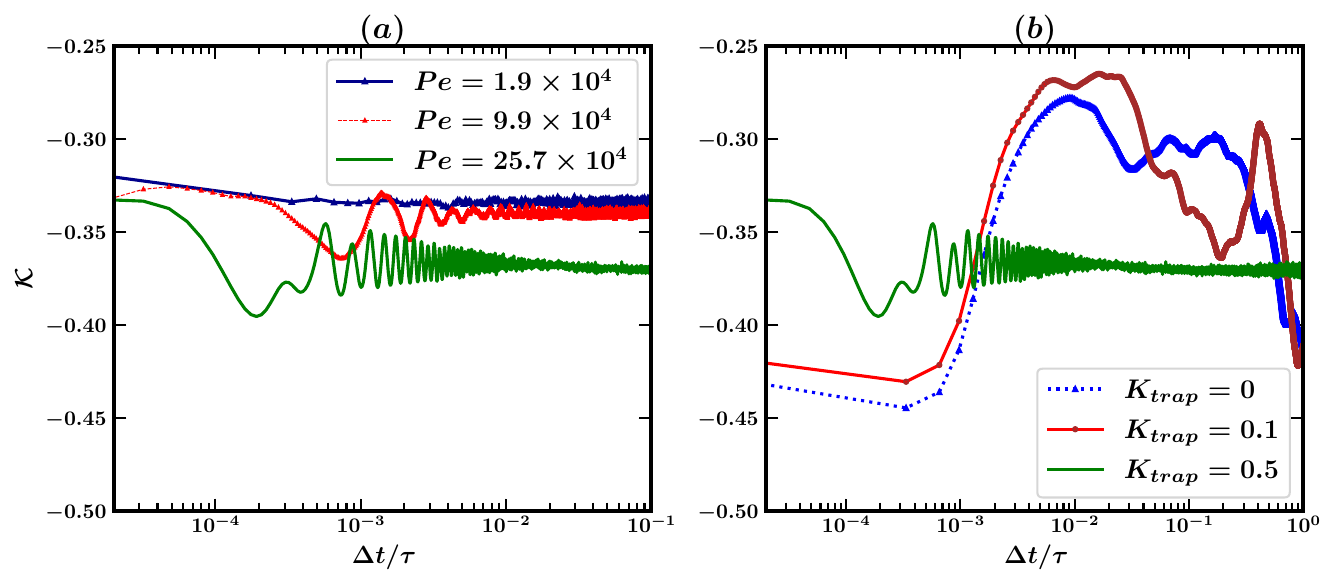}
\caption{\justifying (a) Kurtosis 
$\mathcal{K}$ vs time for varying P\'eclet numbers at fixed trap strength $K_{\text{trap}} = 0.5 ~\KT/\sigma^2$ and bare processivity rate $\Omega=0.5$.
(b) Kurtosis vs time for fixed $Pe = 25.7 \times 10^4$, and varying $K_{\text{trap}}$. Excess kurtosis serves as an additional comparative tool to look at the dual effect of confinement and activity.}
\label{fig7}
\end{figure*}

Introducing a weak trap (\(K_{trap} = 0.1~\KT/\sigma^2\)) (see Fig.~\ref{MSD}(b)) alters the $MSD$ significantly. For \(Pe = 1.9 \times 10^4\), the trap suppresses the long-range center of mass diffusion, with the $MSD$ saturating to a constant value for \(\Delta t/\tau > 10^{-3}\). However, at shorter times (\(\Delta t/\tau < 10^{-3}\)), a diffusive regime is still observed. For moderately higher activity (\(Pe = 9.9 \times 10^4\)), the $MSD$ exhibits diffusive behaviour up to \(\Delta t/\tau \approx 10^{-3}\), followed by saturation accompanied by oscillations, likely induced by the combination of both confinement and activity effects on the polymer~\cite{patel2024exact}. For large activity (\(Pe = 25.7 \times 10^4\)), the polymer escapes the trap, leading to a ballistic regime up to \(\Delta t/\tau\approx 10^{-3}\) before transitioning to diffusive behavior. Note that for a weak trap, there is still a finite probability of escaping the trap at low activities. This might happen very late, and it was not observed within the time scale of our simulations. A stronger trap (\(K_{trap} = 0.5~\KT/\sigma^2\)) suppresses diffusion at longer times, regardless of \(Pe\) (see Fig.~\ref{MSD}(c)). Within the diffusive regime, the $MSD$ increases with \(Pe\), reflecting the enhanced space exploration within the trap facilitated by increased activity. Again, for very high activity, one may expect the polymer to escape the trap.  

We further note the oscillations in the MSD at intermediate $Pe$ values for different trap strengths. These oscillations which have been reported in earlier underdamped systems with confining potentials~\cite{patel2024exact}, arise due to the interplay between active propulsion (quantified by Péclet number $Pe$) and the restoring harmonic confinement.
With higher activity, the amplitude of oscillations decays (see Fig.~\ref{MSD}(c)). The reduced amplitude of oscillations is possibly due to increased polymer interactions with the trap boundary.  


We also examined the autocorrelation of the center of mass orientation, defined as \( C_\theta(t) = \left\langle \exp^{i[\theta(t) - \theta(0)]} \right\rangle \), as shown in Fig.~\ref{cross-cor}(a) for a fixed trap strength and varying $Pe$. Here, $\theta(t)(=\tan^{-1}(y_{\text{cm}}/x_{\text{cm}}))$. For \( K_{\text{trap}} = 0.5\,k_BT/\sigma^2 \), we observe that \( C_\theta(t) \) decays sharply and saturates around \( t/\tau \approx 10^{-3} \) for \( Pe = 1.9 \times 10^4 \). At higher activity (\( Pe = 9.9 \times 10^4 \)), the correlation function decays almost with same rate and  decaying oscillations with larger amplitude emerge around \( t/\tau \approx 10^{-3} \). For even higher activity (\( Pe =25.7 \times 10^4 \)) the correlation function decays sharply and unlike previous case, we see emergence of high frequency oscillations but with smaller magnitude. These oscillations arise due to the rotation of the orientation vector. At these values of $Pe$, spirals form. As these spirals rotate under the active
drive, we see oscillations in the orientation correlation. Note that the oscillations increase in frequency with increasing activity. This is a combination of activity, inducing spiral formation, and circular confinement, further promoting the formation of spirals. 

\subsection{Non-Gaussian statistics of center-of-mass fluctuations}
Although excess kurtosis is a useful non-equilibrium measure used in multiple situations in active matter research~\cite{patel2024exact,pattanayak2024impact,kim2022active,kim2022active,yadav2023dynamics}, due to the finite nature of the confinement in our study, it is expected that the excess kurtosis: $\mathcal{K} = \frac{\langle r_{cm}^4 \rangle}{3\langle r_{cm}^2 \rangle^2} - 1$,
would be non-zero even for $Pe = 0$. However, it is still instructive to look at this measure to see its variation in time as the trap strength and activity are varied.

Figure~\ref{fig7}(a) shows the trend of excess kurtosis of the centre of mass (COM) under a strong confinement potential $K_{\text{trap}} = 0.5~\KT/\sigma^2$. For all nonzero values of $\text{Pe}$, the kurtosis exhibits negative value saturation, signifying non-Gaussian behaviour even for small activity levels, such as $\text{Pe} = 1.9 \times 10^4$. For slightly higher activity ($\text{Pe} = 9.9 \times 10^4$), oscillations appear in an intermediate time regime $10^{-3} \leq \Delta t/\tau \leq 10^{-2}$. The frequency and the amplitude of these oscillations increase with higher activity ($\text{Pe} = 25.7 \times 10^4$). These oscillations result from the interplay between activity and confinement. Confinement leads to stable spiral formation and gliding of the polymer along the edges of the circular trap. Further spirals rotate inside the trap. We see a saturation of $\mathcal{K}$ at late times. The saturation value is again strongly dependent on the activity, with more negative values of kurtosis with increasing $Pe$.

Figure~\ref{fig7}(b) illustrates the effect of the confinement potential on kurtosis for a fixed activity value of $\text{Pe} = 25.7 \times 10^4$. Without confinement, the polymer freely moves on the assay, producing a non-monotonic kurtosis profile. At shorter times, the polymer exhibits gliding motion, while at later times, spiral-like motions dominate, leading to a reduction in kurtosis (less negative). For a weak confinement potential ($K_{\text{trap}} = 0.1~\KT/\sigma^2$), the polymer remains mostly free but occasionally encounters the trap, hindering its motion and affecting the kurtosis. For stronger confinement ($K_{\text{trap}} = 0.5~\KT/\sigma^2$), the polymer becomes fully trapped. Similarly to the previous case, the system transitions to a stationary state through oscillations. This clearly shows the effect of the trap on the non-Gaussian parameter and indicates the critical role played by the two competing factors of confinement and activity.

\section{Conclusions}
In this study, we investigated the dynamics of an actively driven semiflexible polymer confined by a harmonic trap, inspired by motility assays of cytoskeletal filaments propelled by motor proteins. Using a coarse-grained agent-based model incorporating stochastic motor (un)binding and directed forces, we systematically explored the role of activity, polymer stiffness, motor processivity, and confinement strength.

We constructed dynamical behavior maps that reveal the transition from a trapped to a free polymer state as a function of P\'eclet number, trap strength, and processivity. Our results show that polymer flexibility significantly influences confinement: softer polymers remain trapped over a larger range of activity, while stiffer polymers more readily escape. At moderate confinement and activity, the polymer adopts stable spiral conformations, driven by the interplay between active forces, bending elasticity, and excluded volume interactions.

The center-of-mass dynamics exhibited rich behavior across parameter regimes. In the absence of confinement, we observed ballistic-diffusive crossovers characteristic of active polymer systems. The introduction of confinement led to saturation in the mean squared displacement at long times, consistent with restricted motion. We further analyzed the non-Gaussian features of polymer displacement through excess kurtosis, finding that confinement and activity together induce nontrivial  fluctuations of the polymer trajectory.

We also explored how varying the simulation box size, and thereby the available free area outside the trap, affects the probability of the polymer remaining confined. Our results indicate that increasing the free area facilitates escape, particularly at higher activities, while strong confinement consistently promotes trapping across system sizes.

We obtained a scaling form for the critical $Pe$ required for escape of the polymer from the trap. Although the proposed scaling relation for the critical P\'eclet number exhibits reasonable agreement with simulations, it omits contributions from rotational dynamics, polymer stiffness-dependent torque, and fluctuation-driven re-entries. These effects may become non-negligible in regimes of high activity, large $l_p/L$, or under more complex trap topologies, and thus require further theoretical refinement. 

Our theoretical predictions are directly amenable to experimental verification in in-vitro motility assay systems. The viscosity of the cellular environment is approximately $100$ times that of water, with the viscosity of water given by $\eta_w = 0.001$pN-s $\mu m^2$~\cite{howard2002mechanics}. Therefore, the viscosity of the motility assay $\eta = 0.1$pN-s $\mu m^2$, which gives viscous damping over bond-length $\sigma$ as $\gamma = 3\pi\eta\sigma$. Activity of MPs is tuned by ATP concentration. For kinesins, bare MP velocities can range from $0.01 \mu m/s$ to $1 \mu m/s$ when ATP concentrations are varied from $1\mu$M to $1$mM~\cite{schnitzer2000force}. With $k_BT = 4.2\times10^{-3}$pN-nm, and filament length of $10\mu$m, $Pe \approx 2 \times 10^4$ and unit of time $\tau \approx 15.2$ hours. 

Taken together, our study highlights the complex interplay of activity, stiffness, and confinement in determining the dynamical states of motility assay driven polymers. These findings are relevant for experimental setups involving confined motility assays, optical trapping of filaments, and synthetic active matter systems. 
An effective method to confine biofilaments is through optical trapping, which is often used in biological filament motor systems as an essential tool for manipulating and measuring forces. For example, using optical trapping, it is shown that the force generated by a few growing parallel-acting filaments is about $1~\text{pN}$~\cite{footer2007direct}. The method has also been used to trap whole cells~\cite{zhong2013trapping,volpe2006dynamics}. Our predicted transitions from confined to unconfined states and spiral stabilization could be probed using time-resolved tracking of labeled filaments within circular optical traps. Future studies could explore the collective behavior of multiple interacting polymers, as well as the influence of time-dependent or spatially structured confinement landscapes on polymer dynamics—for instance, the extension of microtubules in motility assays, where vortex formation has been observed~\cite{sumino2012large}
\\

\section{Author Contributions}
S.R. and A.K.D. conceived and designed the research. S.R. carried out the simulations, collected the data, and performed the analysis. A.C performed the scaling analysis. All authors contributed to the interpretation of the results and the writing of the manuscript.\\

\section{Conflicts of interest}
There are no conflicts to declare.

\section{Data availability}
The data associated with the figures is available at: \url{https://github.com/AnilBiophysics/Data_Motility_Assay_New}


\bibliography{reference} 
\clearpage
\appendix
\onecolumngrid

\section*{Supplementary Information}

The supplementary material provides additional
figures that support and extend the findings discussed
in the manuscript. Specifically, it includes data on the
dynamical behavior maps of a trapped polymer as the
trap size is varied while the polymer size remains fixed.
Furthermore, it presents turning number distributions
for various bare processivity values.

\begin{figure*}[hb!]
\includegraphics[width=\textwidth]{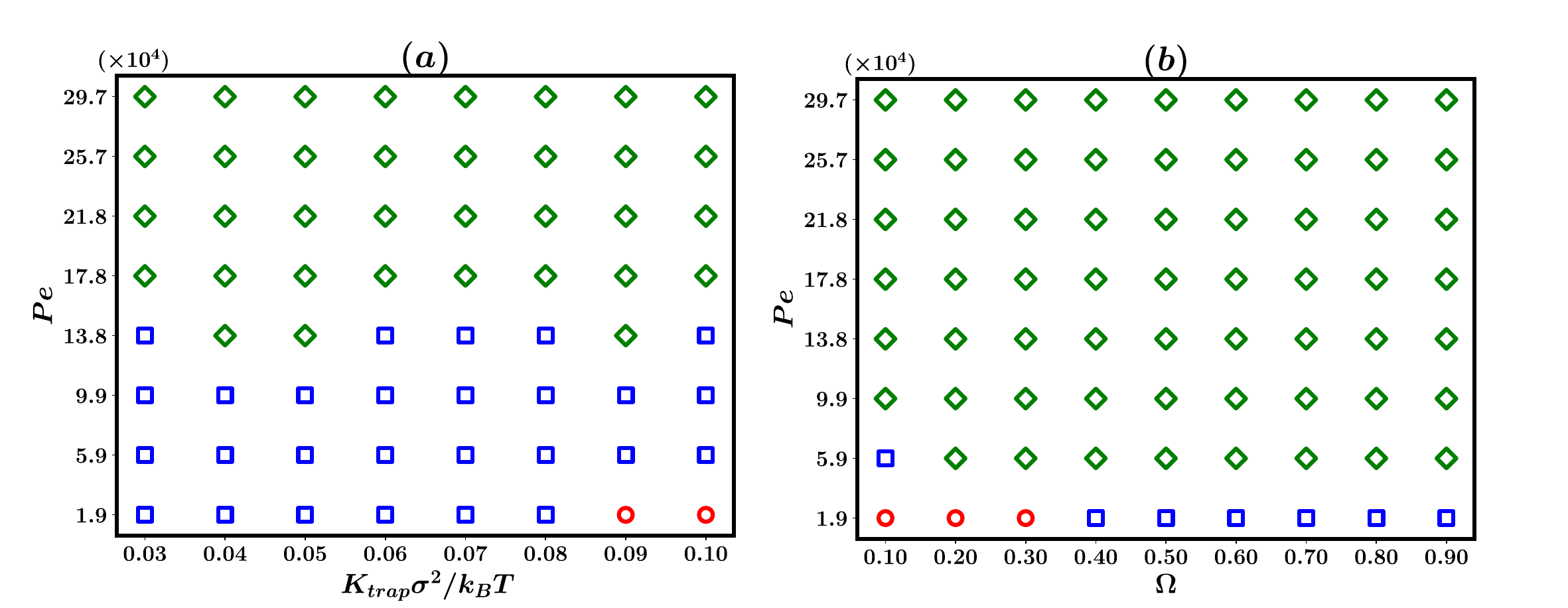}
\caption{\justifying
State diagram for a smaller trap radius. The green diamonds 
({\color{OliveGreen}\Large $\diamond$}) represent free states and the red circles 
({\color{Mahogany}\Large $\circ$}) indicate trapped states. 
Blue squares ({\color{RoyalBlue}$\square$}) represent the co-existence phase. 
The trap radius is $\mathcal{R} = 10\sigma$ and the contour length of the polymer is $L = 63 \sigma$. 
Persistence ratio is $\ell_p/L = 0.30$. 
(a) Bare processivity $\Omega = 0.5$. 
(b) Trap strength $K_{\text{trap}} = 0.08 ~k_B T/\sigma^2$.
}
\label{Fig1}
\end{figure*}

\begin{figure*}[hb!]
\includegraphics[width=\textwidth]{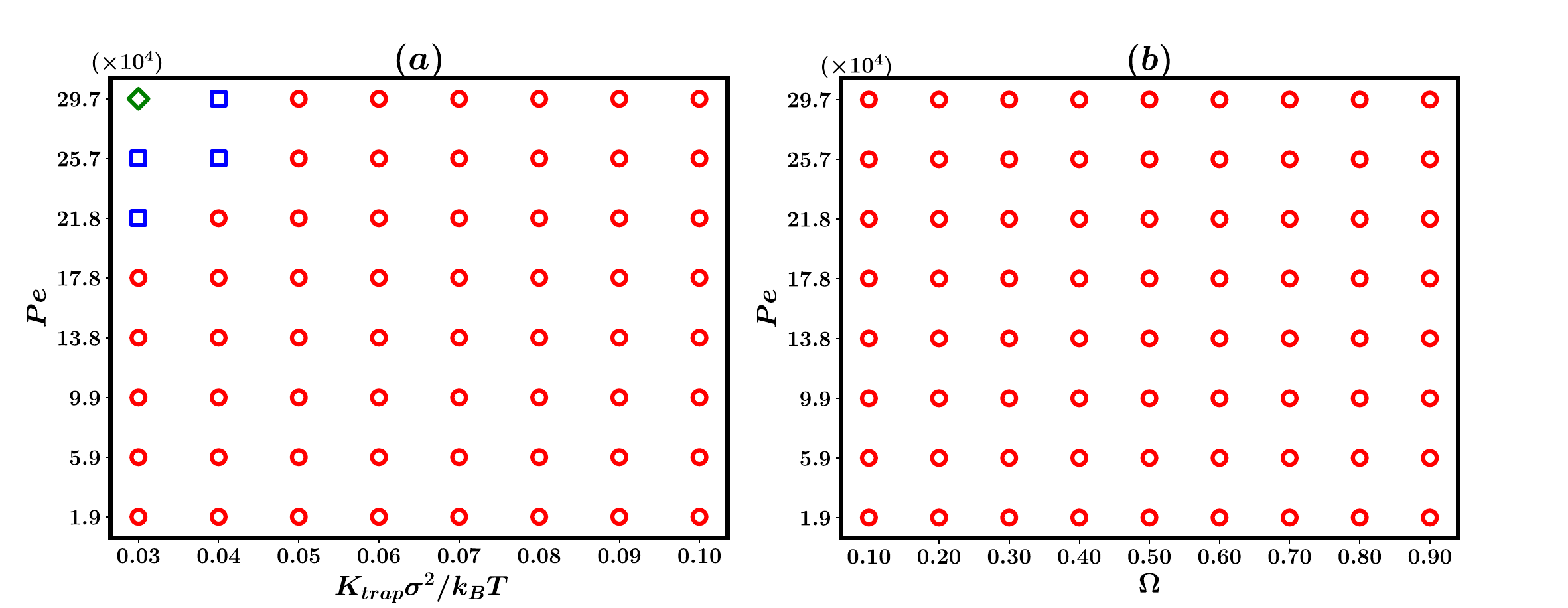}
\caption{\justifying
State diagram for a larger trap radius. Same labeling convention as Fig.~\ref{Fig1}. 
The trap radius is $\mathcal{R} = 50\sigma$ and the contour length of the polymer is $L = 63\sigma$.
}
\label{Fig2}
\end{figure*}

\begin{figure*}[htb!]
\includegraphics[width=\textwidth]{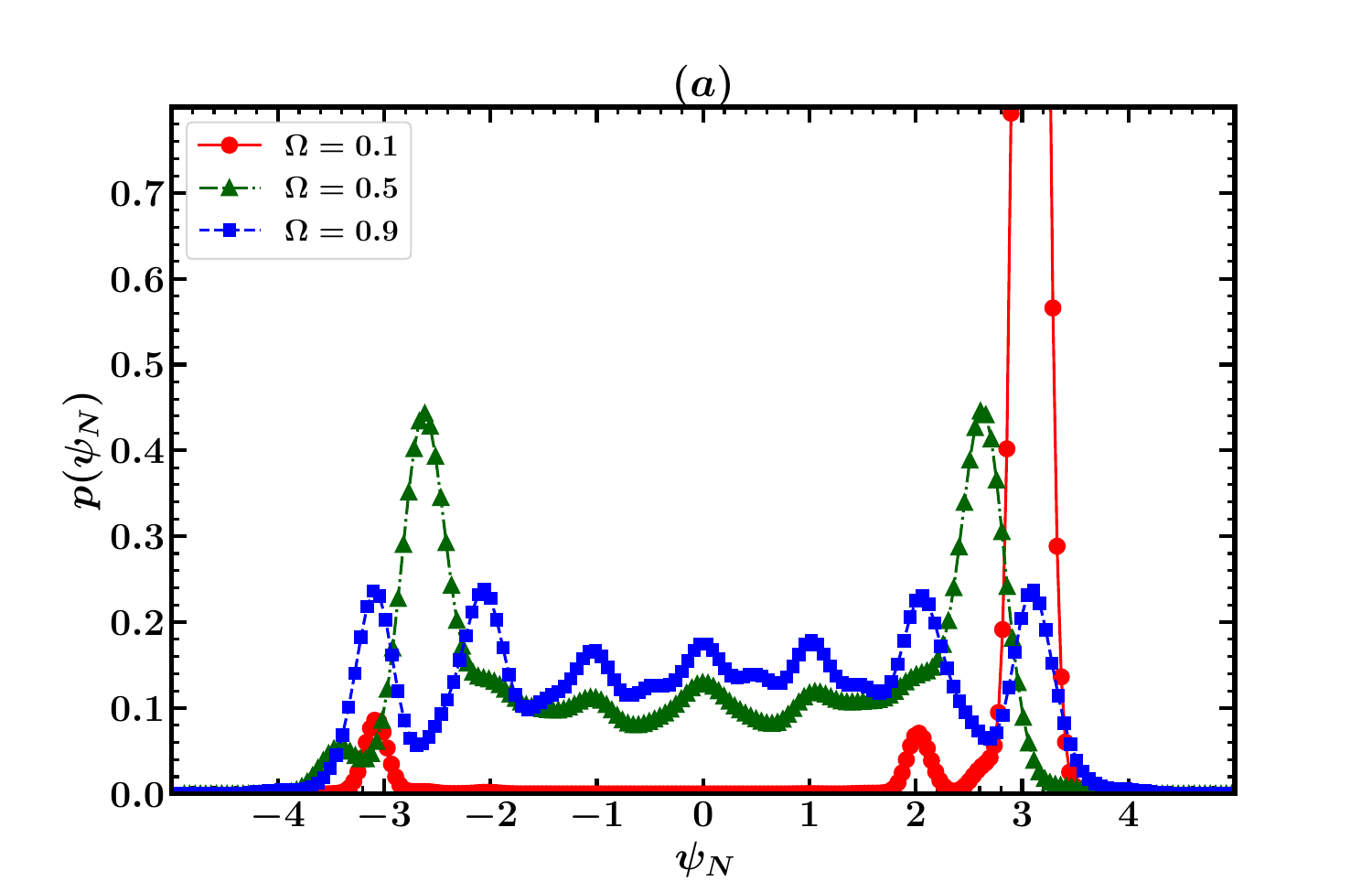}
\caption{\justifying
Steady-state probability distributions of the turning number for 
$\text{Pe} = 9.9 \times 10^4$ at three different values of bare processivity 
$\Omega = 0.1, 0.5, 0.9$ for a trap strength $K_{\text{trap}} = 0.5~k_B T/\sigma^2$.
}
\label{Fig3}
\end{figure*}

\end{document}